\documentclass[aps,prb,letterpaper,superscriptaddress,twocolumn,showpacs,floatfix,10pt]{revtex4-1}
\usepackage{amsfonts}
\usepackage{amsmath}
\usepackage{color}
\usepackage{amssymb}
\usepackage{graphicx}
\usepackage{epstopdf}
\usepackage{epsfig}

\newcommand{\be}[1]{ \begin{eqnarray} \mbox{$\label{#1}$} }

\newcommand{\ee}{\end{eqnarray}}
\newcommand{\pref}[1]{(\ref{#1})}

\begin{document}
\title{Many-Body Localization and Level Repulsion}
\author{Jonas A. Kj\"{a}ll}
\email{jonas.kjall@fysik.su.se}
\affiliation{Deparment of Physics, Stockholm University, Albanova University Center, SE 106 91 Stockholm, Sweden}

\begin{abstract}
Insertion of disorder in thermal interacting quantum systems decreases the amount of level repulsion and can lead to many body localization.
In this paper we use the many body picture to perturbatively study the effect of level repulsion in the localized phase.
We find that most eigenstates can be described accurately in an approximate way, including many with rare resonances.
A classification of the rare resonances shows that most types are exponentially rare and requires exponential fine tuning in an approximate description.
The classification confirms that no rare thermal eigenstates exist in a fully localized phase and we argue that all types of resonances need to become common if a continuous transition into a thermal phase should occur.

\end{abstract}

\maketitle

\section{INTRODUCTION}
Localization in quantum systems prevents transport~\cite{Anderson:1958vr}.
Interacting systems experience \textit{many-body localization} (MBL) at strong enough disorder and do not thermalize~\cite{Basko:2006hh,Gornyi:2005pl,Oganesyan:2007ex}.
However, these many-body systems still have logarithmically slow entanglement growth through long range dephasing saturating at sub-thermal values~\cite{Bardarson:2012gc,Serbyn:2013he}.
Many MBL properties occur in highly excited energy eigenstates and can be accurately described starting from a local picture~\cite{Serbyn:2013lo,Huse:2013li}.
Recently, good progress has been made in developing approximative numerical methods based on matrix product states, to enable the study of larger MBL systems than previous possible~\cite{Pollmann:2016pb,Yu2017:pl,Wahl:2017px}.
Also, experiments with utracold atoms have started to probe MBL physics~\cite{Kondov:2015pl,Schreiber:2015sc,Choi:2016sc,Bordia:2017ax}.
Challenges related to long range behavior, include rare resonances, the phase transition to a thermal phase, especially at the many-body mobility edge~\cite{Basko:2006hh,Monthus:2010gd,Huse:2013bw,Vosk:2014ud,Bauer:2013jw,DeLuca:2013ba,Kjall:2014,Grover:2014ar,Bera:2015pl,Vosk:2015px,Potter:2015px,Imbrie:2016pl,Khemani:2017px,Khemani:2017pl,Pekker:2017pl,Serbyn:2017pb}.
At extensive energies in the many-body spectrum there is significantly less level repulsion in an MBL phase compared to a thermal phase.
In fact, most levels do not repel each other at all and the lack of level repulsion between nearest levels has been used successfully since the first numerical study of MBL~\cite{Oganesyan:2007ex}.
In this paper we take a different approach and investigate the level repulsion a single level in an MBL phase feels from all other levels.
Starting from many-body product states, the eigenstates in the exactly localized limit where no level repulsion is present, we construct a method that perturbatively finds the levels that shift a specific energy level the most. 
A resonance occur, when two levels get so close so a significant mixing of the old eigenstates happens.
Quantities like the entanglement entropy can then change substantially.
We find rare resonances on all length scales, but they become exponentially rarer with increasing distance.
Incorporating those in any approximate description is tricky, since they require exponential fine tuning.
As the transition to the thermal phase is approached the long ranged resonances become more common and we argue it is the proliferation of the longest ones that drive the transition.
While the method developed here can not reach the system sizes needed at the transition, the argument of perturbatively adding more level repulsion suggests a sharp transition as a function of energy. 
\section{THE MODEL}
\begin{figure}[tbp]
  \begin{center}
    \includegraphics[width=85mm]{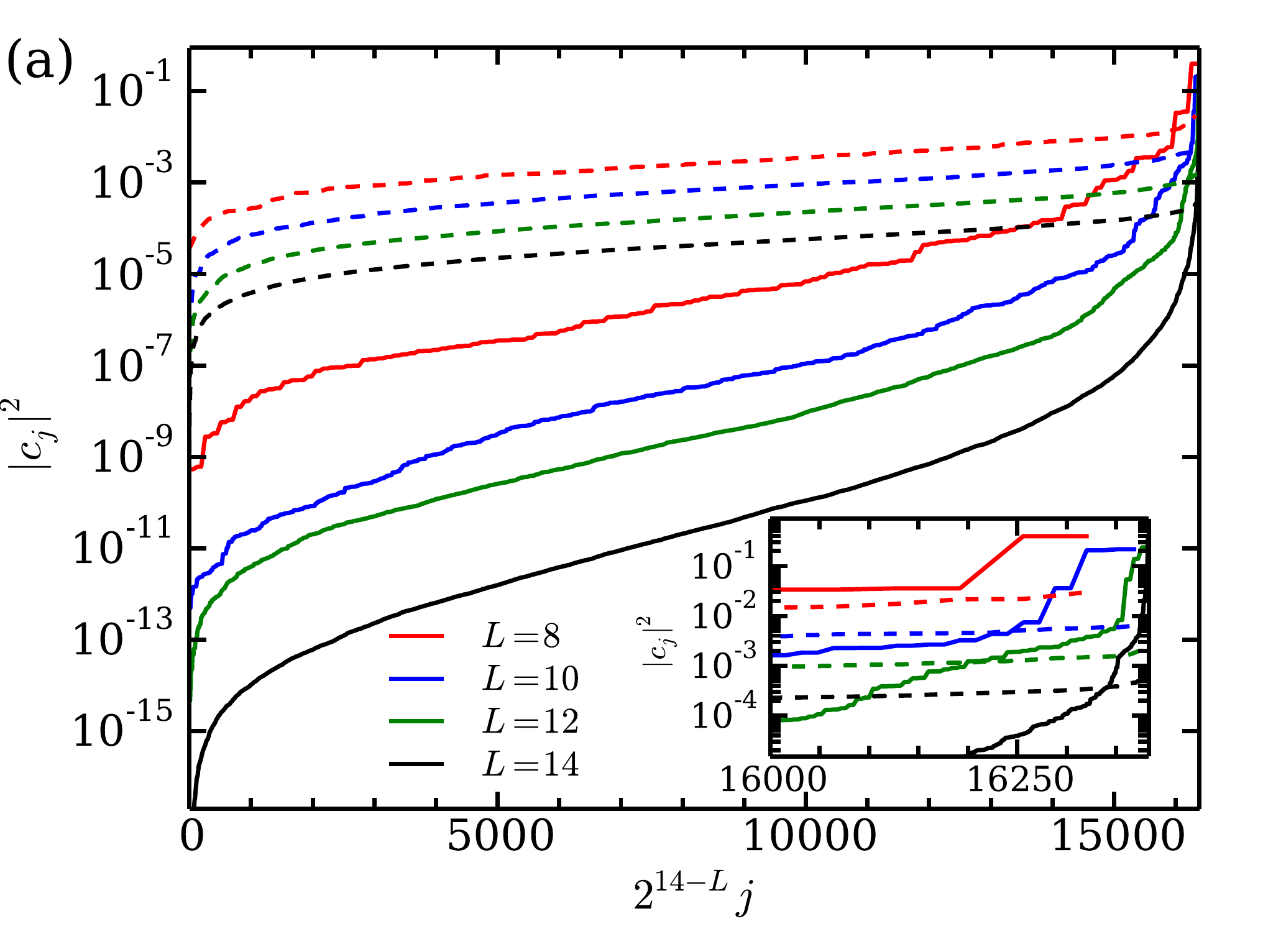}\\
    \includegraphics[width=85mm]{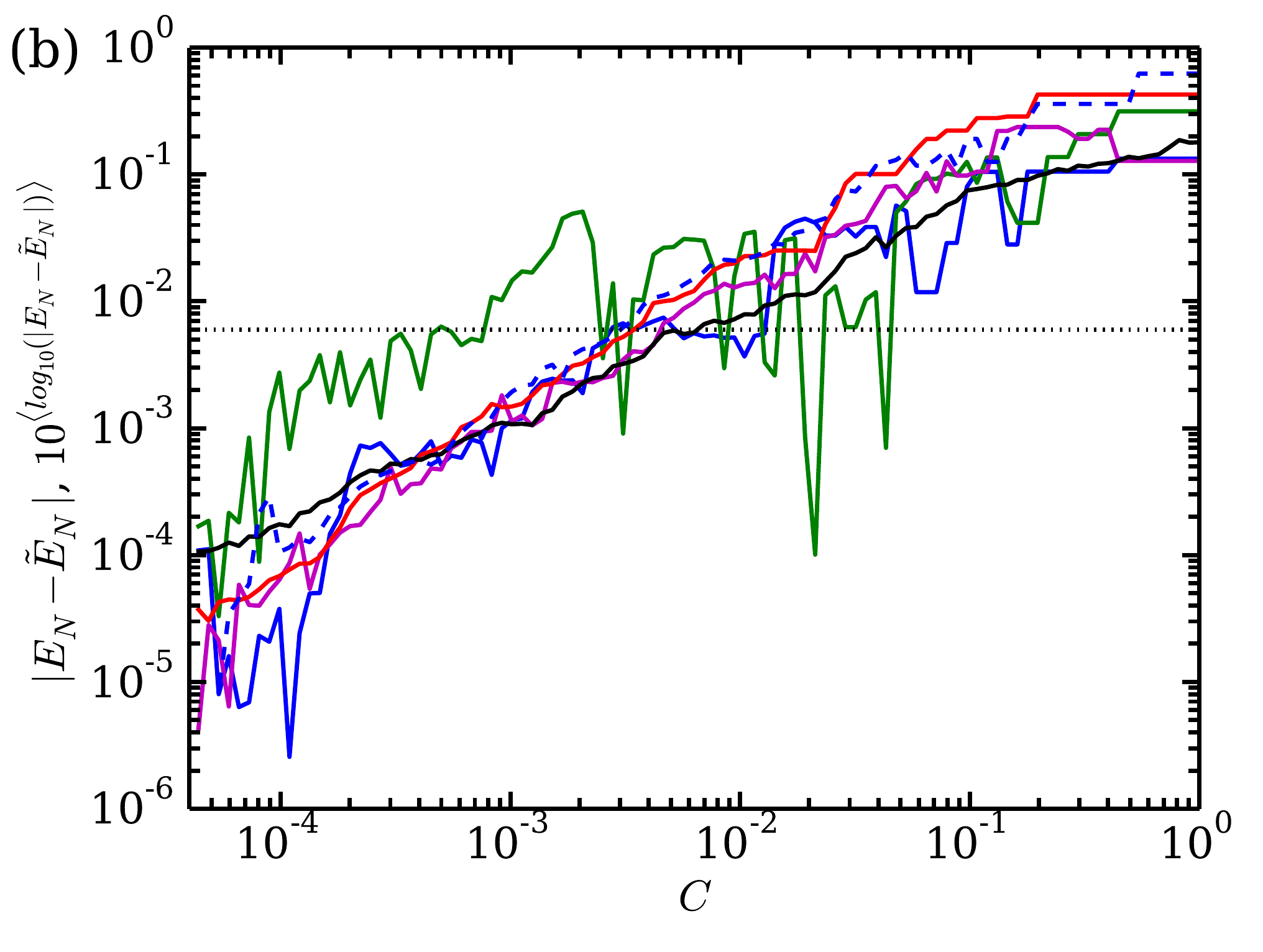}
    \caption{(a) Sorted weights $|c_j|^2$ for random eigenstates in the middle of the spectra for thermal states (dashed lines), from random matrices~\cite{Page:1993da}, and MBL states (solid lines), from Eq.~\pref{eq:H} with $\delta J=6.0$, for system sizes $L=8,10,12,14$ (in descending order), using ED. (b) Energy errors as more states are added to $\tilde{H}_N$ with decreasing $C$ [Eq.~(\ref{eq:cut-off})] at $\delta J=8.0$. Four typical states (color) and a $N_\text{dis}=100$ random realization average (black) for $L=14$ (solid lines). One typical $L=24$ state (dashed line).}
    \label{fig:En}
  \end{center}
\end{figure}

While most of the discussion here is general to all MBL Hamiltonians, for the specifics we consider the transverse field quantum Ising chain with disordered couplings and a next-nearest neighbor Ising term
\begin{equation}
H=-\sum_{a=1}^{L-1}J_a \sigma_a^z\sigma_{a+1}^z + J_{nnn}\sum_{a=1}^{L-2}\sigma_a^z\sigma_{a+2}^z  +  h\sum_{a=1}^{L}\sigma_a^x,
\label{eq:H}
\end{equation}
studied in Ref.~\onlinecite{Kjall:2014}.
Here $\sigma^{x}$ and $\sigma^{z}$ are Pauli matrices and $L$ the number of sites in the chain.
The couplings $J_a=J+\delta J_a$ are independent, with all $\delta J_a$ taken from a uniform random distribution $[-\delta J,\delta J]$.
We set $J=1$, $J_{nnn}=h/2=0.3$ and obtain MBL in all eigenstates for $\delta J\gtrsim 3.8$~\cite{Kjall:2014}.
The  Hamiltonian~(\ref{eq:H}) has a global $\mathbb{Z}_2$ symmetry, and can be written in two blocks (sectors), that both have the same energy spectrum deep in the MBL phase~\cite{Huse:2013bw}.
We split up the Hamiltonian in two parts $H=H_z+H_x$ and treat $H_x=h\sum_{a=1}^{L}\sigma_a^x$ as perturbations. 
$H_z$ is a diagonal matrix, exactly localized since its excitations, domain walls in the ferromagnetic phase, can not move, and its energy eigenvalues $e_j$ are easy to calculate for any system size $L$.
The eigenstates are product states $|j\rangle=|\dots\uparrow\uparrow\downarrow\uparrow\dots\rangle$ of the eigenstates of the $\sigma_a^{z}$ operators, which we number by $j=\sum_a b_a2^{a-1}$ where $b_a=0,1$ if the a$^\text{th}$ spin is down/up.
We write the eigenvalues $E_N$ and the normalized eigenstates 
\be{eq:pES}
|N\rangle=c_n|n\rangle+\displaystyle\sum_{j\neq n} c_j|j\rangle
\ee	
of $H$ in the eigenbasis $|j\rangle$, with $c_j$ constants, keeping the same numbering for $|J\rangle$ as for $|j\rangle$.
In an MBL phase most product states $|j\rangle$ contribute much less to $|N\rangle$ than in a thermal phase as can be seen in Fig.~\ref{fig:En}(a).
The typical weights in thermal states scales as $|c_j|^2\propto 2^{-L}$, the same as the number of basis state increase with system size and hence most of them are needed for an accurate description of the state.
For a typical MBL state it is different, most weights decrease faster than the number of basis state increase with system size.
In the model Eq.~\pref{eq:H}, there are $\binom{L}{p}$ product states that differs from $|n\rangle$ with $p$ spins.
As will be clearer later, all the weights with a finite value of $p$ can in principle be calculated and are finite (albeit often tiny) in an MBL phase.
However, these are just a measure zero compared to the product states of the form $p\approx L/2$ in the thermodynamic limit and their weights are exactly zero.
Consequently, it is reasonable to assume that many properties of each MBL eigenstate in a finite system can be described accurately with a limited amount of product states.
The non-zero weights $c_j$ in $|N\rangle$ can be explained with level repulsion and below we try to find the most relevant ones for a specific eigenstate $|N\rangle$ in an MBL phase using perturbative techniques.
Note, some $|c_j|$ can be larger than $|c_n|$, but in our algorithm we will keep track of which product state $|n\rangle$, that $|N\rangle$ develops from.
\section{LEVEL REPULSION}
In the Hamiltonian $H$, each level is to first order in $h$ connected to $L$ different levels and through them connected in higher orders to the other levels.
We write a two level Hamiltonian as
\begin{equation}
   H_\text{2-level}=\left(
   \begin{array}{cc}
   e+\delta e/2 & t \cr
   t & e-\delta e/2
  \end{array}
  \right),
\end{equation}
where $t$ is the repulsion connecting the two levels and $\delta e$ the energy gap without repulsion. 
Many arguments in this paper will go back to this simple Hamiltonian, using $t$ to different orders in $h$, incorporating the effects from the other levels in $\delta e$ and $t$.
A $p=1$ example is $e_n,e_m=e\pm\delta e/2$ and $h=t$, if two levels $n$ and $m$ that differ by one spin flip are the only ones connected.
As will be discussed, the most important contributions are often from low orders in $h^p$ since $t$ on average decrease exponentially with $p$.
The level repulsion shifts the energy levels
\begin{equation}
  \delta e\rightarrow\sqrt{\delta e^2 + 4t^2},
  \label{ACgap}
\end{equation}
and mixes the weights  in the two (unnormalized) states
\begin{equation}
   \left(
   \begin{array}{c}
   1 \cr
    \frac{-\delta e + \sqrt{\delta e^2 + 4t^2}}{2t} 
  \end{array}
  \right),
\text{and} 
   \left(
   \begin{array}{c}
    -\frac{-\delta e + \sqrt{\delta e^2 + 4t^2}}{2t} \cr
    1
  \end{array}
  \right).
  \label{ACvec}
\end{equation}
The energy shift is larger the larger $t^2/\delta e$ is, but deep in the MBL phase, where the energy gaps get larger (average scale with $\delta J$) and the states are randomly spread out, all shifts are small and the contributions from each level can be treated separately.
\section{PERTURBATIVE MBL}
Here we describe a method that perturbatively tries to find an approximation $\tilde{E}_N$ and $|\tilde{N}\rangle$ to one eigenenergy $E_N$ and one eigenstate $|N\rangle$.
A reduced Hamiltonian $\tilde{H}_N$ containing the diagonal elements $e_j$ with   
\be{eq:cut-off}
\begin{array}{c}
	\text{max}\cr
{\scriptstyle ml\dots k}
\end{array}
\left[\frac{h^{p}}{(h+\delta e_{nm})(h+\delta e_{ml})\dots (h+\delta e_{kj})}\right]\geq C, 
\ee
and the off diagonal Hermitian pairs $h$ connecting them is constructed.
Here, $\delta e_{ml}=e_m-e_l$ is the energy difference between two levels connected by a single spin flip $h$ and $C$ is an arbitrary constant we decrease to include more product states $|j\rangle$ in $|\tilde{N}\rangle$.
After exact diagonalization (ED) of $\tilde{H}_N$ it is only one eigenvalue $\tilde{E}_N$ and one eigenvector $|\tilde{N}\rangle$ we are interested in and we can keep track of which by following the weight structure $\tilde{c}_j$ of $|\tilde{N}\rangle$ with decreasing $C$.
Eq.~(\ref{eq:cut-off}) repeatedly uses first order perturbation theory $h/\delta e_{ml}$, which well describes the typical case of widely separated levels $\delta e_{ml}>>h$ in an MBL phase.
For numerical simplicity, only the maximal contribution from the $p!$ ways the product states $|j\rangle$ and $|n\rangle$ differing by $p$ spin flips can be connected through other product states are used.
The rare cases of levels that repel each other strongly will be discussed later. 
For now, we only modify the factors $h/\delta e_{ml}\rightarrow h/(h+\delta e_{ml})\leq 1$ so we know when to stop searching for more levels $j$ while moving down the tree structure of Eq.~(\ref{eq:cut-off}). 
As the thermal transition is approached, the above prescription does not find the $j$ levels in exactly the correct order as the shifts $e_j\rightarrow E_J$ get larger.
However, it is a random model with shifts as likely to be positive as negative and they always remain reasonable small $|E_J-e_j|<<E_\text{max}-E_\text{GS}$, where $E_\text{max}$ is the maximal energy eigenvalue and $E_\text{GS}$ the ground state energy.
Since, we anyway want a relatively large amount of levels $j$ in $\tilde{H}_N$ it is not essential that they are added in exactly the right order.
The size of $\tilde{H}_N$ sets the numerical limitation of the method and the system sizes $L$ that can be reached depend on the desired accuracy and how deep in the MBL phase the eigenstate is.
Fig.~\ref{fig:En}(b) show $4$ typical examples of $|E_N-\tilde{E}_N|$ (color) and the average $10^{\langle \text{log}_{10}(|E_N-\tilde{E}_N|)\rangle}$ from $N_\text{dis}=100$  disorder realizations (black) as a function of $C$ at $\delta J=8.0$ in a $L=14$ system (solid lines). 
$\langle \cdot\rangle$ indicates averaging over all studied realizations.
The relevant energy scale, the average level spacing in a sector in the middle of the spectrum, is included as reference (black dotted line). 
Larger system sizes behave in the same way.
The blue dashed line is an example from a $L=24$ site system with $E_N$ from linear interpolation. 
The energy error scale linearly with $C$, but individual states can have more or less fluctuations around the average.
More fluctuations occur when the energy levels $\tilde{E}_J$ are shifted significantly more in one direction than in the other.
The approximations made so far are justifiable if no rare resonances are present or if we are just interested in $\tilde{E}_N$ [Eq.~\pref{ACgap} is not as sensitive to large $t$ as Eq.~\pref{ACvec}].
The remainder of this paper investigates these rare resonances in detail and we return to the perturbative MBL algorithm, once we understand them better.

\begin{figure}[tbp]
  \begin{center}
    \includegraphics[width=85mm]{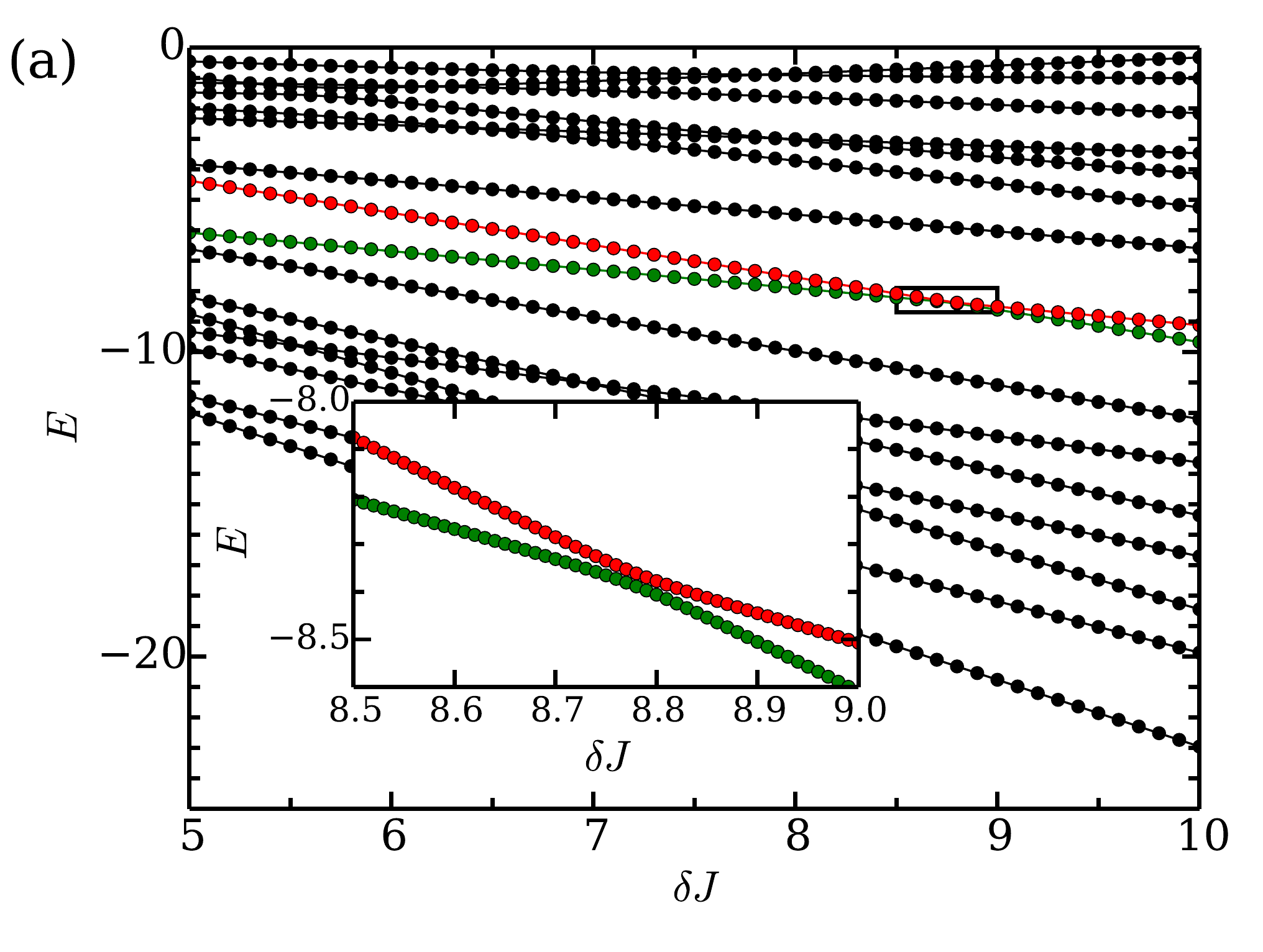}\\ 
    \includegraphics[width=85mm]{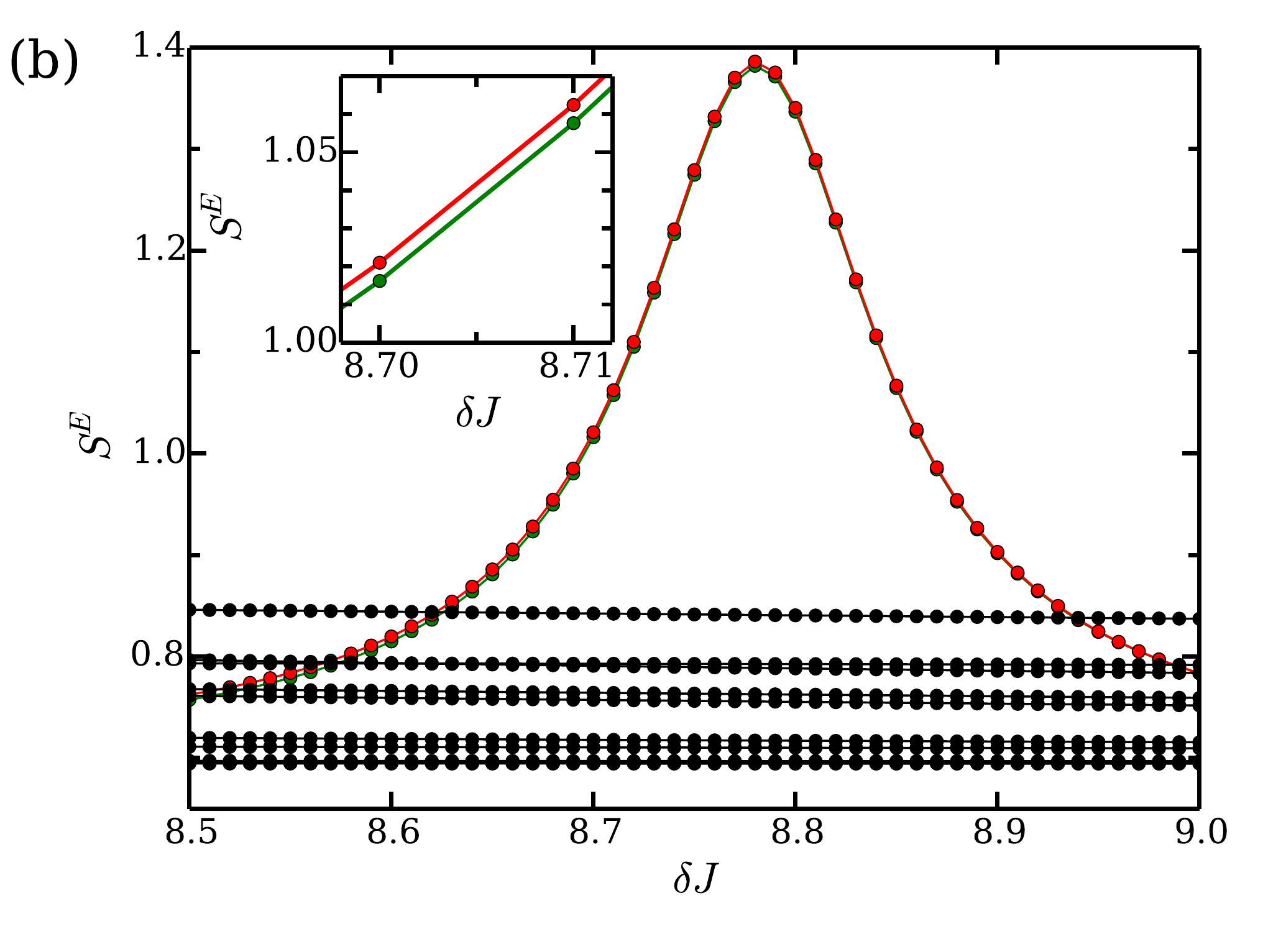}
    \caption{ED. (a) Lower half of the energy spectrum in one sector in a $L=6$ system. Two levels with a rare resonance (inset from box) are highlighted (red and green). (b) The entanglement entropy for the $16$ states in (a). Same $\delta J$ range as the inset in (a).}
    \label{fig:EE}
  \end{center}
\end{figure}

\section{RARE RESONANCES}
\begin{figure}[tbp]
  \begin{center}
    \includegraphics[width=85mm]{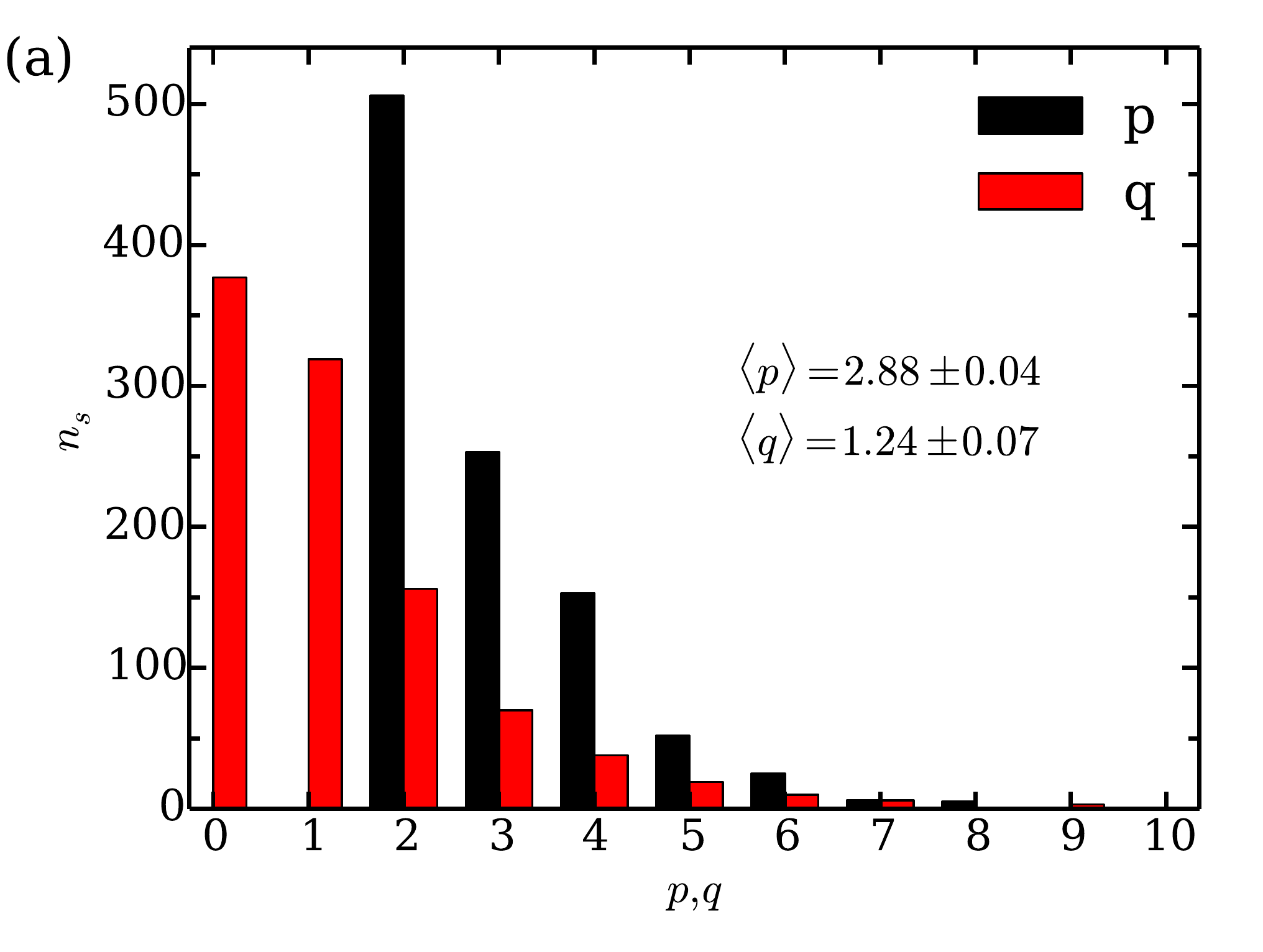}\\
    \includegraphics[width=85mm]{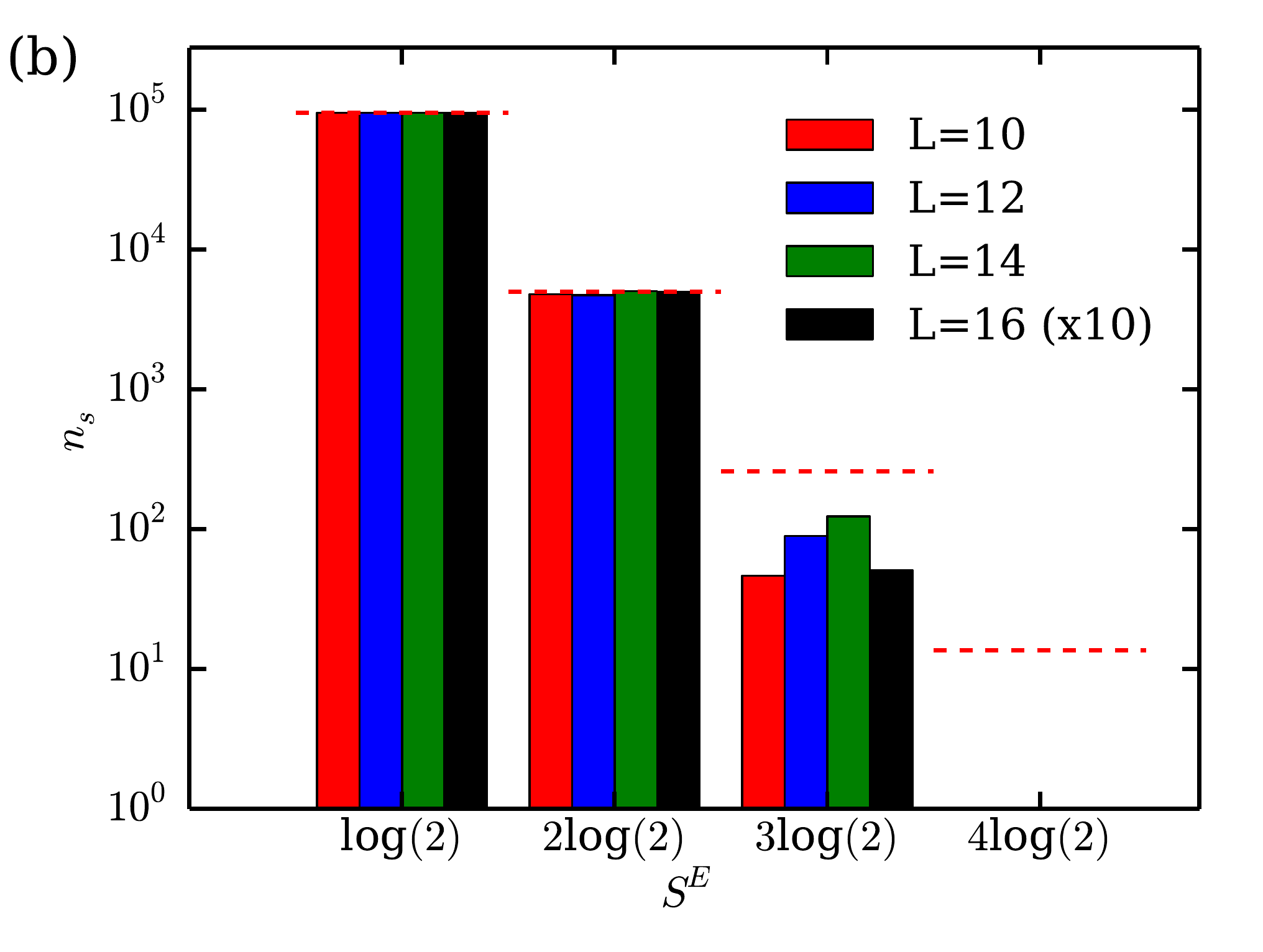}
    \caption{(a) The number of $n_s$ eigenstates with rare resonances labeled after their $p$ (black) and $q$ (red) values, with entanglement entropy $1.5\text{log}(2)<S<2.5\text{log}(2)$. $N_{rr}=1000$ rare resonances detected in a random state  in the middle of the $L=12$ spectrum at $\delta J=8$ from $N_\text{dis}=16256$ disorder realizations with ED. (b) Number of eigenstates $n_s$ with entanglement entropy  $(w-1/2)\text{log}(2)<S_E<(w+1/2)\text{log}(2)$ for $L=10$ (green), $L=12$ (blue), $L=14$ (blue) and $L=16$ (black) from $N_\text{dis}=10^5$ random states ($N_\text{dis}=10^4$ for $L=16$) at $\delta J=10$ with ED.}
    \label{fig:EE2}
  \end{center}
\end{figure}

Rare resonances occur in an MBL spectra between states $|N\rangle$ and $|S\rangle$ far apart in space ($|n\rangle$ and $|s\rangle$ differ by many spin flips) but close in energy $|E_N-E_S|\lesssim t$, see Fig.~\ref{fig:EE}(a) for an example.
The range of a rare resonance in $\delta J$ (or $1/h$) scales linearly with $t$.  
The energy shift is typically small, since $t$ and hence $\delta e$ remain small [Eq.~\pref{ACgap}], and is not particular important for $\tilde{E}_N$. 
However, rare resonances are important in making $|\tilde{N}\rangle$ a good approximation of $|N\rangle$, since $|c_s|\approx|c_n|$ [Eq.~\pref{ACvec}].
An observable that can be very sensitive to resonances, as highlighted in Fig.~\ref{fig:EE}(b), is the von Neumann entanglement entropy
\be{eq:EE}
S_N^E=-\text{Tr}_L\rho \text{ln}\rho,
\ee
with $\rho=\text{Tr}_R|N\rangle\langle N|$ the reduced density matrix for a system cut into a left and a right part at one of the bonds. 
Entanglement entropy obeys an area law in MBL systems and has been a useful quantity in MBL studies, see for example Refs.~\onlinecite{Bauer:2013jw} and~\onlinecite{Kjall:2014}.
A state is entangled across a spatial cut if it can be split up in two parts with different spin configurations on both sides of the cut. 
For two product states, the maximal entanglement entropy is $\text{log}(2)$ if they in addition have equal weights $|c_n|=|c_m|=1/\sqrt{2}$.
In the studied model [Eq.~\pref{eq:H}], deep in the MBL phase, all eigenstates are cat states $|N\rangle,|2^L-1-N\rangle\rightarrow\frac{1}{\sqrt{2}}(|n\rangle\pm|2^L-1-n\rangle)$, between global spin flips and hence have entanglement entropy $S^E_J(\delta J\rightarrow\infty)=\text{log}(2)$.
To get a state with entanglement entropy
\be{eq:EEg}
S^E=w\text{log}(2),
\ee
the least amount of product states $|j\rangle$ needed is $2^w$, if all have equal weights and different spin configurations on both sides of the cut.
If their spin configuration only differ on one side of the cut there is no entanglement entropy increase and it can even decrease if a product state is added that is the same as the previous ones on both sides.
An example of a cut in the middle zero entanglement state, built up of two entangled cat states, is the equal weight state $1/2(|\uparrow\uparrow|\uparrow\uparrow\rangle+|\downarrow\downarrow|\downarrow\downarrow\rangle+|\uparrow\uparrow|\downarrow\downarrow\rangle+|\downarrow\downarrow|\uparrow\uparrow\rangle)$.
Deep in the MBL phase it is easy to get an approximation of how rare different resonances are.
We find the states $|N\rangle$ with an additional entanglement entropy of $\sim\text{log}(2)$ compared to the lowest entangled states and analyze the spin configuration $|j\rangle$ of the largest $|c_j|$ in Eq.~\pref{eq:pES}, that has a spin configuration that differs from $|n\rangle$ (and $|2^L-1-n\rangle$) on both sides of the cut.
Fig.~\ref{fig:EE2}(a) shows the number of rare resonances as a function of $p$, the number of spin flips, and $q=a-b+1-p$, the number of un-flipped spins between the flipped spins, with $a/b$ the position of the left-/right-most flipped spin.
Most of the rare resonances are due to $p=2$ product states, even if the number of potential resonating levels increase with $p$, highlighting that $t$ falls off fast with $p$. 
Also interesting is the $q$-dependence, with most rare resonances from flips of entire domains, followed by flips only separated by an un-flipped spin.
This can be understood studying a $4$-level model 
\begin{equation}
   H_\text{4-level}=\left(
   \begin{array}{cccc}
   e_1 & h & h & 0 \cr
   h & e_2 & 0 & h \cr
   h & 0 & e_3 & h \cr
   0 & h & h & e_4
  \end{array}
  \right),
\end{equation}
with a $p=2$ rare resonance for $e_1\approx e_4$ and $|e_1-e_{2,3}|>>h$.
Importantly, in an approximate 2-level model for levels $1$ and $4$, perturbation theory show to high accuracy $t$ decreasing linearly with $|e_1+e_4-(e_2+e_3)|$, when $|e_1+e_4-(e_2+e_3)|\lesssim|e_1-e_4|$.
In our short range model we have 
\be{3op}
|e_n+e_{n\pm 2^a\pm 2^b}-e_{n\pm 2^a}-e_{n\pm 2^b}|= 
\begin{cases}
4J_a \quad \text{ if } b-a=1\\
0\text{ otherwise }\\
4J_{nnn} \text{ if } b-a=2
\end{cases}
\ee
with $b>a$ and the signs $\pm$ determined by the configuration $|n\rangle$.
With added kinetics (nonzero $h$), two well separated spins/domains (large $q$), in an MBL phase still interact exponentially weakly and the energy cost of flipping one is next to independent of flipping the other.
Numerically (see also below) we indeed find an exponential decrease 
\be{texp}
\langle t\rangle &\sim\langle |E_N+E_{N\pm 2^a\pm 2^b}-E_{N\pm 2^a}-E_{N\pm 2^b}|\rangle\nonumber\\
&\propto e^{-q/\xi},
\label{eq:texp}
\ee
with $q(p=2)=b-a-1$ and $\xi(\delta J,1/h,E/L)$ the relevant length scale for resonances (see below).
Individual shifts $e_j\rightarrow E_J$ are of a much larger magnitude.

All possible resonance types in an MBL phase can be detected in a finite system, since their exponential fall off is sufficiently fast. 
The probability $P_{w-1}$ of a state with entanglement entropy $~w\text{log}(2)$, with $w$ an integer, is independent of $L$ (for large enough $L$) in an MBL phase see Fig.~\ref{fig:EE2}(b) and falls off faster than exponential with $w$ (red dashed lines).
In a random spectrum, rare states with higher entanglement entropy should have occurred with probability $P_{w-1}\sim P_1^{2^{w-1}-1}$ if all resonance types had the same probability and always increased the entanglement.
However, since the probability for high $(p,q)$ resonances decreases exponentially and they do not always increase the entanglement, there is no rare states with thermal entanglement entropy in a full MBL phase.
Full here means the MBL phase extends to all energy densities.
\section{PERTURBATIVE MBL WITH RARE RESONANCES}
Having classified the different types of rare resonances we turn back to our perturbative MBL algorithm.
We check for possible rare resonances $|E_N-E_S|\lesssim t$ and treat them with a form of degenerate perturbation theory.
First, we find $\tilde{E}_S$ to the same accuracy $C$ as $\tilde{E}_N$.
Then, we diagonalize $\tilde{H}_{NS}$ containing all the levels $j$ that build up $\tilde{H}_S$ and $\tilde{H}_N$.
We take it as a resonance if $|\tilde{c}_n|+|\tilde{c}_s|>\frac{5}{4}\text{max}(|\tilde{c}_j|)$, with $\frac{5}{4}$ chosen from numerical tests.
Redo this for all possible rare resonances and the final $|\tilde{N}\rangle$ is obtained by diagonalizing $\tilde{H}_{NS_1\cdots S_N}$.
We construct an algorithm, using $|E_{S_i}-\tilde{E}_{S_i}|\sim C$, that relatively efficient finds the possible rare resonances.
Note, in addition to $|\tilde{N}\rangle$ we also get $|\tilde{S_i}\rangle$ to the same accuracy $C$, but we now have to diagonalize a matrix with up to $i+1$ times as many product states.
Also note, if we are interested in for example $\tilde{S}_N^E$, it is enough to check for resonances across one bond, while for a good approximation of $|\tilde{N}\rangle$ resonances across every bond need to be considered.
\begin{figure}[tbp]
  \begin{center}
    \includegraphics[width=85mm]{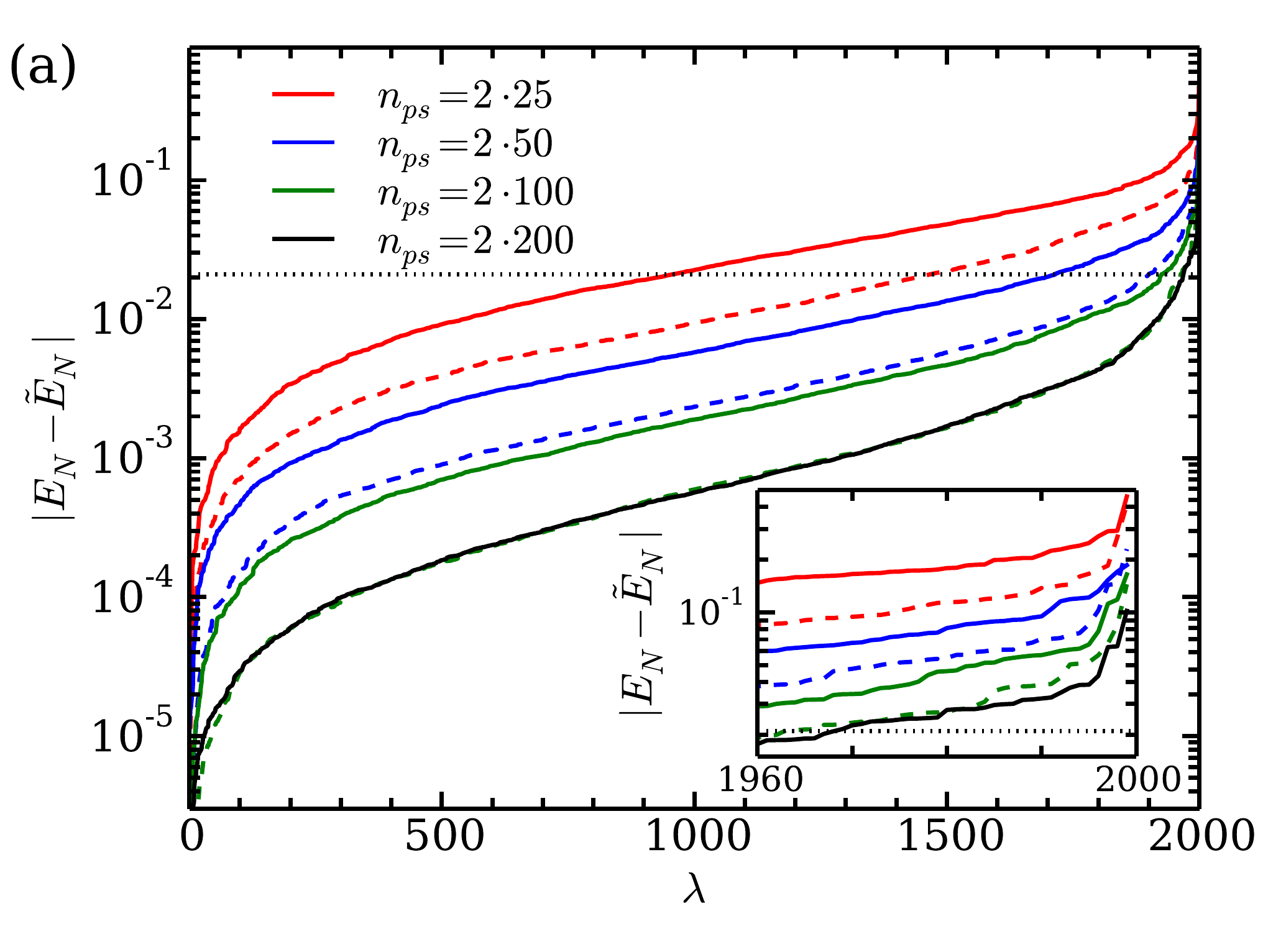}\\
    \includegraphics[width=85mm]{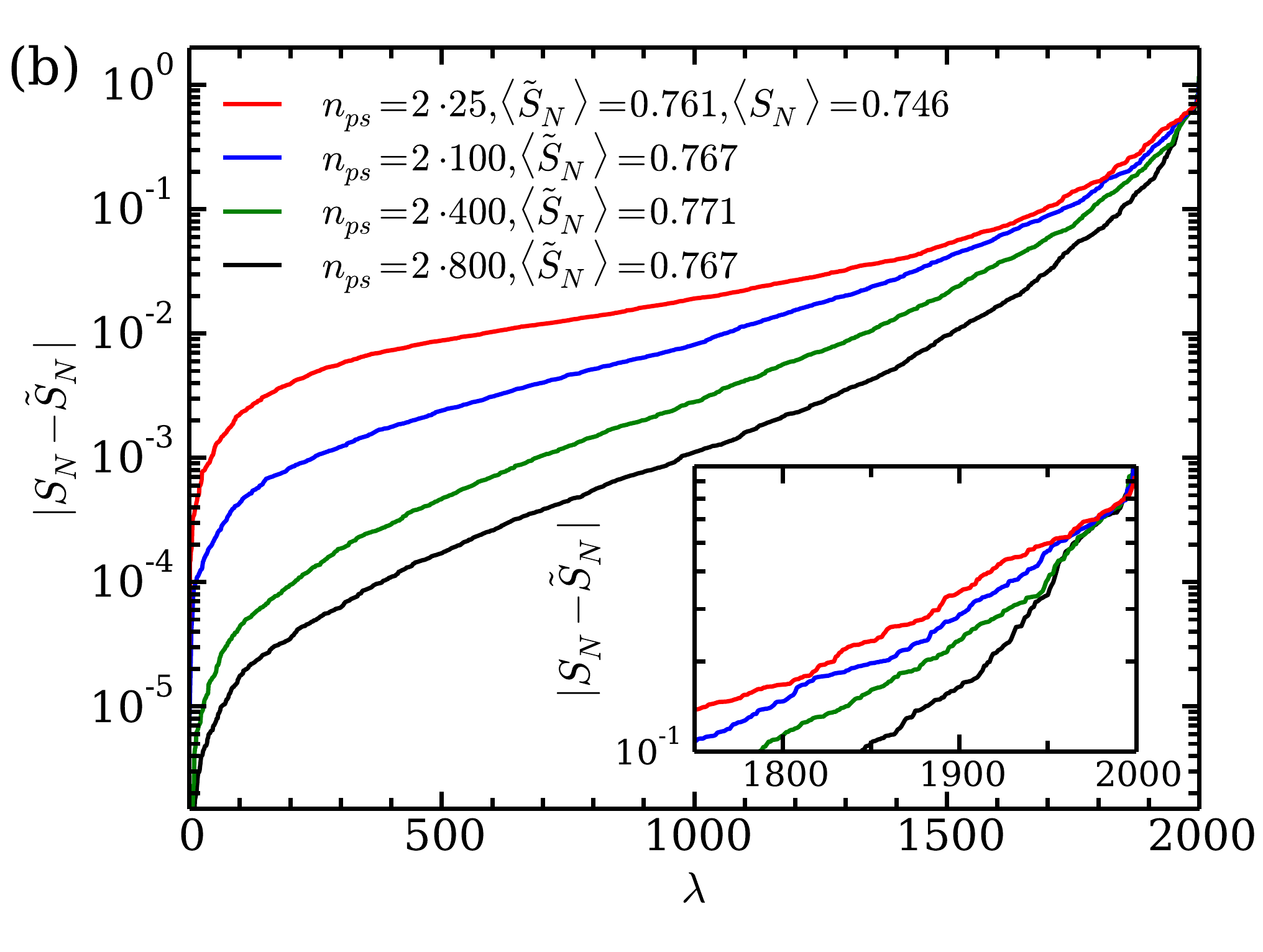}
    \caption{Sorted energy errors $|E_N-\tilde{E}_N|$ (a) and entanglement entropy errors $|S_N-\tilde{S}_N|$ (b) at $\delta J=8$ from $N_\text{dis}=2000$ random disorder realizations $\lambda$ of $\tilde{H}_N$ with $n_{ps}$ product states for $L=12$ (solid lines) and $L=10$ (dashed lines). The average level spacing $\delta E$ in the middle of the spectrum of a $L=12$ system (black dotted line). Insets highlight the largest parts.}
    \label{fig:Res}
  \end{center}
\end{figure}

With a complete perturbative MBL algorithm let us investigate its applicability and limitations.
The energy error $|E_N-\tilde{E}_N|$ normally converges fast with $C$ or the number of product states $n_{ps}$ included in $\tilde{H}_{N}$, see Fig.~\ref{fig:Res}(a), but fluctuations observed in Fig.~\ref{fig:En}(b) can give a slower convergence.
Since we often are interested in average quantities, this is not a problem.
If one is interested in a specific state $|\tilde{N}\rangle$, a look on $\tilde{E}_N$ as a function of $C$ (or $n_{ps}$) give a good idea of how much it fluctuates.
The $2$ in $n_{ps}=2\cdot x$ comes from the inclusions of the global spin flips $|2^L-1-j\rangle$ to keep the symmetry of the studied model.
To find the correct rare resonances one needs $|E_N-\tilde{E}_N|\lesssim t$.
Numerically, we find that $|S^E_N-\tilde{S}^E_N|$ converges well in most states, see Fig.~\ref{fig:Res}(b) while others have extra or missing resonances as expected.
The tail of states with $|S^E_N-\tilde{S}^E_N|\approx\text{log}(2)$ disappear slowly with $C$ and some remain until the full $H$ is diagonalized.
The approximate averages $\langle \tilde{S}^E_N\rangle$ are somewhat higher than expected.
In a random spectrum one could have expected on average to obtain equally many resonances at $\tilde{E}_N$ as at $E_N$. 
This is true for the low $p$ and $q$ resonances, but we find more high $q$ resonances at $\tilde{E}_N$ than at $E_N$. 
The reason is that the small values of $|E_N+E_{N\pm 2^a\pm 2^b}-E_{N\pm 2^a}-E_{N\pm 2^b}|$ at large $q$ are really fine tuned and they are typically larger for larger $|E_N-\tilde{E}_N|$.
There is also a risk of underestimating the fraction of high $p$ resonances at $\tilde{E}_N$ if one does not make sure there is enough of $|j\rangle$ states to connect them to $|n\rangle$.   
The entanglement of resonating states are tricky to analyze with approximate method, but with knowledge of its shortcomings, useful information can still be gained.
\section{PHASE TRANSITION}
\begin{figure}[tbp]
  \begin{center}
    \includegraphics[width=85mm]{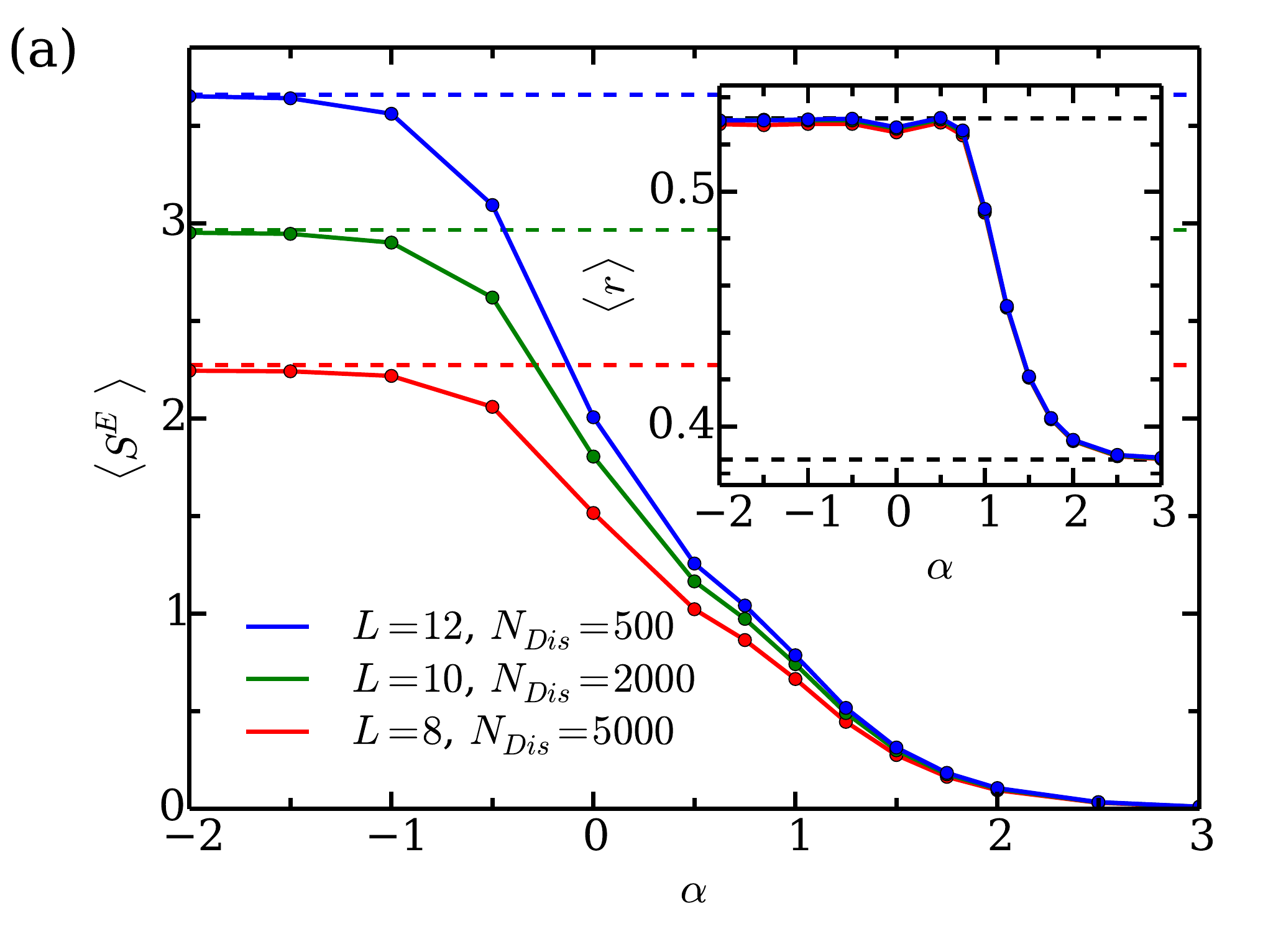}\\
    \includegraphics[width=85mm]{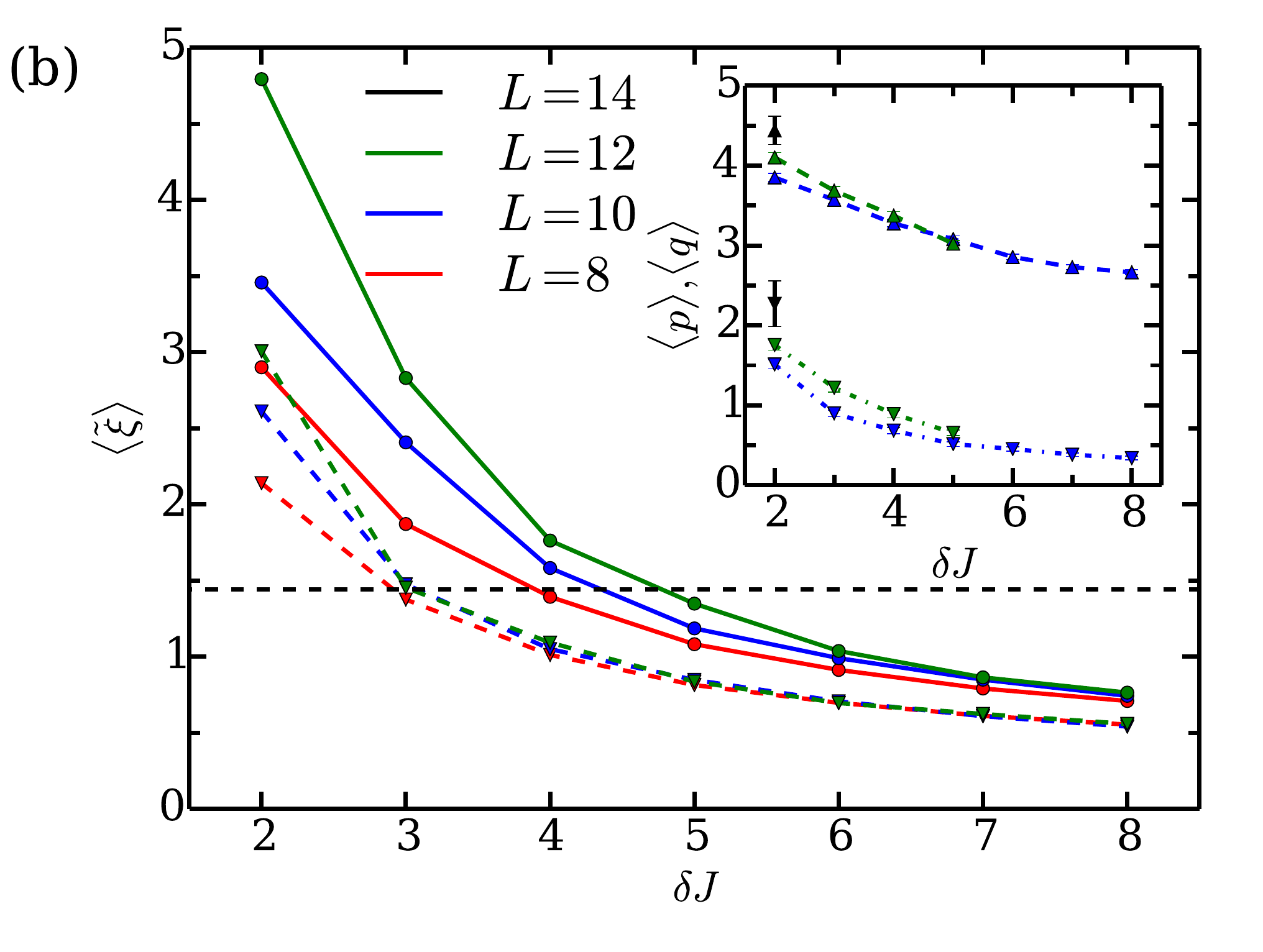}
    \caption{(a) Disorder averaged entanglement entropy $S_E$ and level statistics $r$ (inset) for the thermal Hamiltonian $H_\text{th}$ [Eq.~(\ref{Hth})] as a function of the strength of the level repulsion $t/\langle\delta e\rangle\propto 10^{-\alpha}$ for system sizes $L=8,10,12$ with standard error (not visible) from ED. Dashed lines indicate expected values in the thermal [Eq.~(\ref{eq:Sth}) and $\langle r\rangle_\text{th}=0.531$] and MBL ($\langle r\rangle_\text{MBL}=0.386$) phases~\cite{Oganesyan:2007ex}.   (b) The resonance length scale $\langle\tilde{\xi}\rangle$ [see Eq.~(\ref{eq:texp})] as a function of $\delta J$ in the middle of the energy spectrum (solid lines) and at $E_{1/5}=\frac{1}{2}(E_{GS}-E_\text{max})/5$ (dashed lines).  Dashed line indicates the phase transition $\xi_c$. Data from ED averaged over $N_\text{dis}=10^4$ disorder realizations for $L=8,10,12$.  (Inset) $\langle p\rangle$ (dashed lines) and $\langle q\rangle$ (dashed dotted lines) from single states at $E_{1/5}$ averaged over $N_{rr}=1000$ rare resonances for $L=10$(blue), $12$(green) and $N_{rr}=100$ for $L=14$(black), with standard error from ED.}
    \label{fig:Trans}
  \end{center}
\end{figure}

Next, we discuss what occurs when the phase transition out of the MBL phase is approached. 
As $\delta J$ decreases the average energy level spacing $\langle\delta E\rangle$ get smaller.
On average the shifts $e_j\rightarrow E_J$ increase and $\tilde{E}_N(C)$ fluctuates more. 
More importantly though is that the probability for resonances increases, since they drive the phase transition out of the MBL phase.
Resonating levels $|E_N-E_S|\lesssim t$ share most basis states $|j\rangle$ with weights of the same order of magnitude.
A thermal state obeys a volume law and has entanglement entropy 
\be{eq:Sth}
S^E_\text{th}=\frac{1}{2}[L\text{log}(2)-1]
\ee
at infinite temperature (middle of the energy spectrum) across a cut in the middle of the state~\cite{Page:1993da}.
Hence, if the transition is continuous in the entanglement entropy we need to couple together more and more product states up to basically all $2^{L/2}$ possible on each side of the cut [see Eq.~\pref{eq:EEg}].
Numerically we find that states containing $\gtrsim 2^{0.51L}$ random product states, all with the same weights, have $S^E= S^E_\text{th}$ for large $L$ (not shown).
Since, not all weights are the same at the transition, some more levels are needed in practice, but the minimum number of thermal states needed for a thermal phase $\sim O(2^{L/2})<<O(2^L)$ is much smaller than the number of available states.
To model the thermal phase [of Eq.~\ref{eq:H}] at infinite temperature in the thermodynamic limit we construct a system size independent toy Hamiltonian 
\be{Hth}
H_\text{th}=\left(
   \begin{array}{cccc}
   d_{11} & t_{12} & \cdots & t_{1M} \cr
   t_{21} & d_{22} & \cdots & t_{2M} \cr
   \vdots & \vdots & \ddots & \vdots \cr
   t_{M1} & t_{M2} & \cdots & d_{MM}
  \end{array}
  \right).
\ee
Here $M=2^L$ and $d_{jj}$ are uniformly distributed random numbers between $0$ and $1$, to get an average energy gap $\langle \delta E\rangle=2^{-L}$.
The off-diagonal elements are $t_{jk}=f(1+|\tau_{jk}|)2^{-L}$, with $\tau_{jk}$ a normal distributed random number with mean $0$ and variance $1$ and $f=10^{-\alpha}$ the free parameter we tune.
Most $t_{jk}$'s have similar magnitudes, since most levels differ by $\sim L/2$ spin flips and we expect the level repulsion between nearest levels to dominate.
The precise form of the randomness does not appear to matter, but some randomness in the off-diagonal elements is needed.
We diagonalize $H_\text{th}$ and calculate the level statistics $\langle r\rangle$ and the entanglement entropy $\langle S^E\rangle$, see Fig.~\ref{fig:Trans}(a).
The level statistics is defined as $r=\text{min}(\delta E_{\eta+1}, \delta E_\eta)/\text{max}(\delta E_{\eta+1}, \delta E_\eta)$, where $\delta E_\eta=E_{\eta+1}-E_\eta$ is the energy gap between two nearest energy eigenvalues, see Ref.~\onlinecite{Oganesyan:2007ex} for details.
A system size independent quantity like $\langle r\rangle$ show bascially no system size dependence (the three lines in the plot are on top of each other), except for some energy spectrum edge effects, and $\langle S^E\rangle$ gets very close to $S^E_\text{th}$ in the thermal phase, which occur for $f\gtrsim 10$.
For $f\lesssim 10$ it is no longer a good model of Eq.~\ref{eq:H} since most of its off diagonal elements goes to zero in the MBL phase.
If the phase transition in Eq.~(\ref{eq:H}) is continuous, we expect the resonances between levels differing by $p,q\sim L/2$ to dominate as the transition is approached, since that is what most level differ by.
In Fig.~\ref{fig:EE2}(a), we see that the exponential decrease is simliar in $p$ and $q$.
Assuming it is the same [Eq.~(\ref{eq:texp})], we get the phase transition in the thermodynamic limit $L\rightarrow \infty$ when
\be{eq:Trans}
\langle \delta E \rangle=\langle \delta e \rangle \propto 2^{-L} = e^{-L/\xi_c} \propto \langle t_c(p,q=L/2)\rangle,
\ee 
or when $\xi_c=1/\text{log}(2)$.
Note, that while $e^{-L\text{log}(2)}$ decrease faster with $L$ than the number of resonances with $p,q=L/2$, which is $\binom{L-2}{(L-2)/2}$, increase, it does not decrease faster than the number of resonances with $p,q\sim L/2$ compared to the number of resonances with $p\sim 2$ and $q\sim 0$.
An interesting Ising type toy Hamiltonian for the MBL-thermal phase transition is
\be{eq:Luitz}
H_\text{tr}=\sigma_{-1}^z+\sigma_{-1}^x+\displaystyle\sum_{a=0}^{L-2}\frac{h_a}{2}\sigma_a^z+ \displaystyle\sum_{a=0}^{L-2}\frac{\gamma^a}{4}\sigma_{a}^x\sigma_{-1}^x,
\ee
motivated by Ref.~\onlinecite{Luitz:2017pl}, with $h_a$ taken from a random box distribution $[-0.5,0.5]$.
In this model all spins (but one) are next nearest neighbors and we can directly calculate the level repulsion between two levels differing by two spin flips $t(p=2)\propto\gamma^{n+m}$ (using the Ising duality transformation).
A continuous phase transition, needs to be driven by resonances between levels differing by $\sim L/2$ spin flips.
If $t(p,q=L/2)=\text{min}[t(p=2)]$ also in this model, we get a continuous phase transition at $2^{-L}=\gamma_c^{2L-5} $ when nearest levels resonate or $\gamma_c=1/\sqrt{2}$.
This is in agreement with the numerical results obtained with a thermal region $\text{R}$ in Ref.~\onlinecite{Luitz:2017pl}.
The replacement of a thermal region with a single spin $\text{R}\rightarrow \sigma_{-1}^z+\sigma_{-1}^x$, which need to be interacting to make it an interacting model, is discussed in Ref.~\onlinecite{Ponte:2017ax}.
A continuous phase transition allow for a phase transition as a function of energy, a many-body mobility edge~\cite{Basko:2006hh}.
All eigenstates above will be thermal with extensive entanglement entropy and all below will be localized with finite entanglement entropy.
To get a thermal eigenstate just above the mobility edge there need to be level repulsion between basically all spin configurations on one side of a cut, as in $H_\text{th}$.
However, the level it develops from (upon turning on $h$ for example) will only be resonating $|e_n-e_m|\lesssim t$ with a few other levels, $\sim 10$ according to Fig.~\ref{fig:Trans}(a).
The rest of the weights from levels further away can be thought of as coming from resonating chains of levels, where the levels a level is resonating with is in their turn resonating with other levels and so on.
Just below the many-body mobility edge, the MBL eigenstates develops from levels that can be part of a few resonating chains that goes into the thermal phase.
We emphasize that just being part of a resonating chain is not sufficient for a level to thermalize, since weights from far away in the chain will be too small for extended entanglement entropy, see Eq.~(\ref{eq:EEg}).
A non-zero level repulsion with those levels is also necessary.
However, in an MBL phase where $\xi<1/\text{log}(2)$, a levels level repulsion with most other levels is $t=0$, including with those in a possible nearby thermal phase.
The energy density dependence of $\xi(E/L)$ can for example be seen in Eq.~(\ref{eq:texp}), where the energy difference depends on the amount of level repulsion the involved levels experience, which is strongly correlated with the average energy gap $\langle \delta E\rangle$, which increase continuously with decreasing $E$.
Note, $\langle \delta E\rangle$ is directly present in the condition for the phase transition [Eq.~(\ref{eq:Trans})], but only through its exponential scaling which does not change with $E$, as opposed to $\xi$.
We conclude our discussion of the phase transition with some supporting data.
The exponential decrease in Eq.~(\ref{eq:texp}) can be calculated with ED for any $|E_1-E_4|$ and we get a good approximation $\tilde{\xi}$ to $\xi$ by doing a linear fit to the data at $q\sim L/2$ in a log plot, see Fig.~\ref{fig:Trans}(b).
In the thermodynamic limit $\xi$ is not defined in the thermal phase, but for small finite system we can calculate it.
The transition $\xi_c=1/\text{log}(2)$ is reached for noitceable larger, but still reasonable, disorder strengths compared to $\delta J_c\approx 3.8$ in Ref.~\onlinecite{Kjall:2014} in the middle of the energy spectrum.
A clear energy dependence in $\xi$ is also observed, with the transition at $E_{1/5}=\frac{1}{2}(E_{GS}-E_\text{max})/5$ occuring at a smaller $\delta J$.
To further support the existence of a mobility edge, we numerically investigate $\langle p\rangle$ and $\langle q\rangle$, obtained as in Fig.~\ref{fig:EE2}(a), as a function of $\delta J$ at an energy density $E_{1/5}$, see inset in Fig.~\ref{fig:Trans}(b).
This far down in the energy spectrum, resonances are rarer and we can go to smaller $\delta J$.
Since this approach gets more uncertain, due to more multi-level resonances closer to the phase transition, we stop at $\delta J=2$, where the probability for single resonances is $P_1\sim 0.3$ for the studied system sizes.
Exponential decrease in $p$ and $q$ is observed for all data points.
Closer to the transition the probability for resonances with higher $p$ and $q$ increases, as we argued above was necessary for the transition.
With growing tails the averages $\langle p\rangle$ and $\langle q\rangle$ increase somewhat with system size, but remains well under the thermal value $L/2$.
Note, $\delta J=2$ is well under the calculated phase transition at larger energy densities $\delta J_c\approx 3.8$~\cite{Kjall:2014}.
Apart from the approximations done, also note that this model [Eq.~(\ref{eq:H})] is not in the scaling regime at the phase transition for the systems sizes reachable with ED (see Ref.~\onlinecite{Kjall:2014}).
\section{DISCUSSION}
In this paper we used the low amount of level repulsion to construct a perturbative method for MBL eigenstates.
It is likely more advanced perturbative algorithms than Eq.~(\ref{eq:cut-off}), using perturbation theory to higher orders or using iteratively updated $\tilde{E}_J$ instead of $e_j$, can find the levels $j$ in a better order.
However, they will come with a higher computational cost.
We tried some with little improvement, but more research is needed.
If one keeps the restriction of only treating two levels differing by a spin flip at a time, as in Eq.~(\ref{eq:cut-off}), the full level repulsion expression of Eq.~(\ref{ACvec}), which also has a natural maximum of $1$, can be used instead $h/(h+\delta e_{kl}) \rightarrow (-\delta e_{kl} + \sqrt{\delta e_{kl}^2 + 4h^2})/2h\leq 1$. 
In the common case of well separated levels $\delta e_{lk}>>h$ both versions approaches $h/\delta e_{lk}$.
Rare resonances are important and our detailed study show their probability decrease exponentially with distance.
This observation show unambiguously that rare thermal states can not occur in a full MBL phase.
The exponential decay in Eq.~(\ref{eq:texp}) is probably hard to get correct in any approximative method, since it needs to be fine-tuned, but it would be interesting to investigate the probability for long range (large $p$) resonances in some of the more successful. 
For example, Fig. 5 in Ref.~\onlinecite{Yu2017:pl} appear to also slightly overestimate $\langle\tilde{S}^E\rangle$ in the MBL phase in a similar way as we found in our Fig.~\ref{fig:Res}(b).
Closer to the phase transtion into the thermal phase the probability for resonances increases and if the entanglement entropy is continuous across the transition the long ranged resonances need to become common and the transition should become sharp in energy, a mobility edge.

\acknowledgements
We acknowledge insightful discussions with Frank Pollmann, David Huse, Ehud Altman, Eddy Ardonne and in particular Jens Bardarson. The research has been funded by the Swedish Research Council and some of the initial simulations were done at Max Planck Institute for the Physics of Complex Systems.
%

%

\end{document}